\documentclass[pra, twocolumn , showpacs
			     , showkeys
				         	]{revtex4}
\usepackage{amsmath}
\usepackage{amssymb}
\usepackage{bm}
\usepackage{dsfont}
\usepackage{graphicx}
\usepackage{subfigure}
\usepackage{dcolumn}
\usepackage[normalem]{ulem}
\usepackage{color}

\newcommand{\ri}{\mathrm{i}}
\newcommand{\re}{\mathrm{e}}
\newcommand{\rd}{\mathrm{d}}

\newcommand{\psia}{\hat{\psi}^{\dag}}
\newcommand{\psio}{\hat{\psi}}

\newcommand{\Ho}{\hat{H}}

\newcommand{\rhoo}{\hat{\rho}}

\newcommand{\la}{\langle}
\newcommand{\ra}{\rangle}
\newcommand{\be}{\begin{equation}}
\newcommand{\ee}{\end{equation}}
\newcommand{\bes}{\begin{eqnarray}}
\newcommand{\ees}{\end{eqnarray}}

\newcommand{\br}{{\bm r}}

\newcommand{\bs}{{\bm s}}

\newcommand{\bn}{{\bm n}}
\newcommand{\bB}{{\bm B}}
\newcommand{\bw}{{\bm w}}

\newcommand{\dw}{\downarrow}
\newcommand{\up}{\uparrow}

\newcommand{\tr}{\text{tr}}

\begin{document}

\title{Spin segregation via dynamically induced long-range interaction in a
system of ultracold fermions}
\author{Ulrich Ebling$^1$}
\author{Andr\'e~Eckardt$^{1,2}$}
\email{eckardt@pks.mpg.de}
\author{Maciej~Lewenstein$^{1,3}$}
\affiliation{$^1$ICFO-Institut de Ci\`encies Fot\`oniques, 
Av.\ Canal Ol\'impic s/n, E-08860 Castelldefels (Barcelona), Spain,}
\affiliation{$^2$Max-Planck-Institut f\"ur Physik komplexer Systeme, 
N\"othnitzer Str.\ 38, D-01187 Dresden, Germany}
\affiliation{$^3$ICREA-Institucio Catalana de Recerca i Estudis Avan\c{c}ats,
Lluis Companys 23, E-08010 Barcelona, Spain}
\date{October 6, 2011}

\begin{abstract}
We investigate theoretically the time evolution of a one-dimensional
system of spin-1/2 fermions in a harmonic trap after, initially,
a spiral spin configuration far-from equilibrium is created. We predict a spin
segregation building up in time already for weak interaction under realistic
experimental conditions. The effect relies on the interplay between exchange
interaction and the harmonic trap, and it is found for a wide range of
parameters. It can be understood as a consequence of an effective, dynamically
induced long-range interaction that is derived by integrating out the rapid
oscillatory dynamics in the trap. 
\end{abstract}
\pacs{67.30.hj, 67.10.Hk, 05.20.Dd}
\keywords{Fermi gas, off-equilibrium, exchange interaction, spin
segregation, dynamically induced long-range interaction}
\maketitle

\section{Introduction}

Ultracold atomic quantum gases have been established as a clean and tunable test ground for many-body
physics \cite{BlochEtAl08}. They allow to mimic condensed matter, but offer also opportunities to
study aspects of many-body physics that are hard to address in other systems. An important example
for the latter is the broad and widely unexplored subject of many-body {dynamics}. 
In this paper we investigate theoretically the spinor dynamics of a Fermi gas
far away from equilibrium. We consider a one-dimensional system of repulsively
interacting spin-1/2 fermions confined in a harmonic trap. The initial state is
created out of the spin-polarized equilibrium by rotating the spins spatially
into a spiral configuration (as previously done for a Bose condensate
\cite{Vengalattore08} and proposed for strongly interacting fermions 
for the purpose of probing the Stoner transition \cite{ConduitAltman10}). We
show that already weak interaction, like in the experiment described in Refs.\
\cite{DuEtAl,DuEtAlTheo}, is sufficient to induce a robust spin segregation. The
effect builds up on times that are long compared to the oscillatory motion of
the atoms in the trap. It can be explained as a consequence of an effective,
dynamically created long-range interaction that we obtain by integrating out the
rapid oscillatory dynamics in the trap. Within the framework of a
semiclassical theory, the effective interaction is isotropic in
phase space. 
The fact that away from equilibrium already weak interaction can cause a
noticeable spin segregation contrasts with the equilibrium physics of the
system. For example, the spin segregation of itinerant ferromagnetism, possible
signatures of which have recently been observed in a cold-atom system
\cite{JoEtAl09}, requires strong interparticle repulsion as well as higher
dimensions. The experiment \cite{JoEtAl09} has inspired also theoretical work on
the dynamics of strongly-interacting spin-1/2 fermions, e.g.\ Refs.\
\cite{ConduitAltman10,dyn}.
In this paper we stick, however, to the regime of weak interaction, where the
fermionic cold-atom system does not suffer from effects originating from the
coupling to molecular two-body bound states such as dissipative particle losses
\cite{PekkerEtAl11} or non-universal scattering properties \cite{ZhouEtAl11}.

In the following two sections we will first introduce the system
and describe the semiclassical mean-field theory that we use to simulate its
dynamics. In section \ref{sec:simulation} we then present our numerical
results, predicting a dynamical spin segregation. By integrating out rapid
oscillations in the trap, in section \ref{sec:effective} we derive an effective
description for the dynamics that explains this finding in an intuitive fashion
as a consequence of a dynamically induced long-range interaction. Before we
close with conclusions, experimental signatures are discussed. 

\section{\label{sec:system}System}
We consider a gas of fermionic atoms of mass $M$ having two relevant 
internal states, $m=\frac{1}{2},-\frac{1}{2}\equiv\up,\dw$.
The gas is not necessarily quantum degenerate but sufficiently cold and dilute to
interact via low-energy $s$-wave scattering only. Consequently, the interaction
between two particles is captured by a pseudo
potential 
$g' P_{m_1'm_1,m_2'm_2}\delta(\br_1-\br_2)$
with $(m_1,m_2)$ and $(m_1',m_2')$ denoting the 
spin state before and after scattering, respectively. Here
	\bes\label{eq:projector}
	P_{m_1'm_1,m_2'm_2}
&=& \frac{1}{2}(e_{m_1'm_1}e_{m_2'm_2}-e_{m_2'm_1}e_{m_1'm_2}) 
	\nonumber\\
	&=& \frac{1}{4} e_{m_1'm_1}e_{m_2'm_2}-\bs_{m_1'm_1}\cdot\bs_{m_2'm_2}
	\ees
with unity matrix $e_{m'm}$ and vector of spin-1/2 matrices $\bs_{m'm}$,
projects onto the antisymmetric spin singlet state
two scattering particles have (due
to Fermi statistics and the symmetric $s$-wave state).  The term
$-\frac{1}{2}e_{m_2'm_1}e_{m_1'm_2} =
-\frac{1}{4}e_{m_1'm_1}e_{m_2'm_2}-\bs_{m_1'm_1}\cdot\bs_{m_2'm_2} $,
corresponding to the so-called exchange interaction, gives rise to spin-spin
coupling.
The coupling constant $g'=4\pi\hbar^2a_s/M$ is proportional to the singlet
s-wave scattering length $a_s$, characterizing the actual
interatomic potential. With this, we can write down the Hamiltonaian 
  $
	\Ho' = \int\!\rd\br\, \psia_{m'}(\br)h'_{m'm}(\br)\psio_{m}(\br)
		+ \frac{g'}{2}\int\!\rd\br\, 
		\psia_{m_1'}(\br)\psia_{m_2'}(\br)
		P_{m_1'm_1,m_2'm_2}\psio_{m_2}(\br)\psio_{m_1}(\br)			
  $.
Repeated spin indices imply summation, $\psio_{m}(\br)$ is a fermionic field
operator, and $h'_{m'm}(\br) =-\frac{\hbar^2}{2M}\nabla_\br^2e_{m'm} +
V_{m'm}'(\br)$ denotes the single-particle Hamiltonian containing the
potential $V_{m'm}'(\br)=V(\br)e_{m'm}+\bB(\br)\cdot\bs_{m'm}$ that can be
decomposed into a spin-independent term $V'(\br)$ and an effective magnetic
field $\bB(\br)$ acting on the spin. 

We are interested in the regime where the dynamics is reduced to one spatial
dimension and consider a harmonic confining potential $V'(\br)\equiv 
V(x)+V_\perp(y,z)=\frac{1}{2}M[\omega^2x^2 + \omega_\perp^2 (y^2+z^2)]$ with the
a tight transversal confinement $\omega_\perp$ being large compared to other
relevant energy scales such as the initial temperature or chemical
potential. Therefore, the particles basically occupy
the transversal single-particle ground state. Moreover, we assume that the
magnetic field varies in $x$-direction only, 
$\bB(\br)=B_0{\bm e}_z+\bB(x)$. A possibly present constant part $B_0{\bm e}_z$ of the
magnetic field will be dropped in the following, i.e.\ we are working in a spin frame
rotating around the $z$-axis. Introducing a dimensionless description
in units of the longitudinal trap [energies, lengths, momenta, and times are
given from now on in multiples of $\hbar\omega$, $(M\omega/\hbar)^{\!-1/2}$,
$(M\hbar\omega)^{\!1/2}$, and $\omega^{-1}$, respectively] we arrive at the
Hamiltonian 
  \bes\label{eq:ham}
    \Ho &=& \int\!\rd x\, \psia_{m'}(x)h_{m'm}(x) \psio_{m}(x)
	\\\nonumber&&
		+ \frac{g}{2}\int\!\rd x\, 
		\psia_{m_1'}(x)\psia_{m_2'}(x)
		P_{m_1'm_1,m_2'm_2}\psio_{m_2}(x)\psio_{m_1}(x)
  \ees 
for the one-dimensional problem. Here $g=2\hbar\omega_\perp
a_s\times(M\omega/\hbar)^{\!1/2}/(\hbar\omega)$, 
  \be
    h_{m'm}(x)=  -\frac{1}{2}\partial^2_x e_{m'm} + V_{m'm}(x),
  \ee
and
      \be
	V_{m'm}(x)=\frac{1}{2}x^2e_{m'm}+\bB(x)\cdot\bs_{m'm}.
      \ee
By swapping field operators and indices 
($\psia_{m_1'}\psia_{m_2'}=-\psia_{m_2'}\psia_{m_1'}\to-\psia_{m_1'} \psia_{m_2'}$)
the interaction can be simplified to 
$\Ho_\text{int} = g \int\!\rd x\,
\psia_{\up}(x)\psio_{\up}(x)\psia_{\dw}(x) \psio_{\dw}(x)$ reflecting
that by Pauli exclusion only fermions of opposite spin interact. The resulting
spin coupling -- parallel spins avoid repulsion -- is expressed more clearly,
however, in Eqs.~(\ref{eq:projector}) and (\ref{eq:ham}).

\section{\label{sec:theory}Semiclassical description}

\subsection{Equations of motion}
We study the system's dynamics in terms of the single-particle density matrix 
  \be
    n_{m'm}(x',x) 
  \equiv\tr\{\rhoo \psia_{m'}(x')\psio_m(x)\}
  \ee
with density operator $\rhoo$. 
It evolves in time according to $\ri\dot\rhoo=[\Ho,\rhoo]$, giving 
  \be
  \ri\,\dot n_{m'm}(x',x) 
  =\la[\psia_{m'}(x')\psio_m(x),\Ho]\ra.
  \ee 
using cyclic permutation under the trace. 
While for non-interacting particles the r.h.s.\ of this equation
reads $h_{mk}(x)n_{m'k}(x',x)-h_{km'}(x)n_{km}(x',x)$, the interaction
$\Ho_\text{int}$ leads to quartic expectation values that we decompose like
\be
\la\psia_k\psia_l\psio_m\psio_n\ra
\approx \la\psia_k\psio_n\ra\la\psia_l\psio_m\ra
-\la\psia_k\psio_m\ra\la\psia_l\psio_n\ra
\ee
in order to get a closed equation 
for $n_{m'm}(x',x)$. By Wick's theorem this decomposition is exact for the
initial state considered here, being an equilibrium state of a quadratic
Hamiltonian modified only by the spiral spin rotation generated by another
quadratic Hamiltonian. At later times it corresponds to the time-dependent
Hartree-Fock approximation that is suitable for weak interaction and leads to
the non-linear equation of motion
	\be\label{eq:motion}
	\ri\dot n_{m'm}(x',x) 
		= h^\text{mf}_{mk}(x) n_{m'k}(x',x)-h^\text{mf}_{km'}(x')n_{km}(x',x).
	\ee
The Hartree-Fock Hamiltonian $h^\text{mf}_{m'm}(x) = h_{m'm}(x)+
V_{m'm}^\text{mf}(x)$ 
comprises the mean-field potential
$V^\text{mf}_{m'm}(x)=V_\text{mf}(x)e_{m'm}+\bB_\text{mf}(x)\cdot\bs_{m'm}$ 
where
	\bes\label{eq:mf}
	V_\text{mf}(x)&=& \frac{1}{2}gn_0(x),
	  \nonumber\\
	\bB_\text{mf}(x) &=&-2g\bn(x),
	\ees
with particle density $n_0(x)=e_{m'm}n_{m'm}(x,x)$ and spin density
$\bn(x)=\bs_{m'm}n_{m'm}(x,x)$. 

In a next step, we introduce the Wigner function
	\be
	w_{m'm}(x,p) = \frac{1}{2\pi}\int\!\rd \xi\, 	
					\re^{-\ri p\xi}n_{m'm}(x-\xi/2,x+\xi/2).
	\ee
Using Eq.~(\ref{eq:motion}), this phase-space representation of the 
single-particle density matrix can be shown to evolve like (e.g.\
\cite{Schleich})
	\bes\label{eq:series}
	\dot w_{m'm}(x,p)=-p\partial_x w_{m'm}(x,p)
			+\frac{1}{\ri}\sum_{\alpha=0}^\infty\frac{1}{\alpha!}
			\Big(\frac{\ri}{2}\partial_y\partial_p\Big)^{\!\alpha}
	\nonumber\\	\times
	[\bar V_{mk}(y)w_{m'k}(x,p)
			-(-)^\alpha \bar V_{km'}(y)w_{km}(x,p)]\Big|_{y=x}
	\ees
with $\bar V_{m'm}(x) 
	    \equiv V_{m'm}(x)+V_{m'm}^\text{mf}(x)$.
The relation
      \be
      n_{m'm}(x,x)=\int\!\rd p\,w_{m'm}(x,p)
      \ee
connecting the spatial densities entering the mean-field potential to the Wigner function 
closes this equation of motion. 
We employ a semiclassical approximation to the \emph{motional} degrees of freedom by
truncating the infinite series after $\alpha=1$. This is justified as long
as the potential $\bar V_{m'm}(x)$
varies slowly compared to the single-particle wave lengths involved. It is, thus,
particularly suitable for sufficiently hot or dense systems, with either the thermal 
or the Fermi wave length 
small. Moreover, the truncation is exact for harmonic or linear potentials 
$\bar V_{m'm}(x)$. It gives 
\bes 
	\dot w_{m'm}
		&=&-p\partial_x w_{m'm}
			+ (\partial_x\bar V)\partial_pw_{m'm}
	  \nonumber\\&&
		-\ri\bar\bB\cdot(\bs_{mk}w_{m'k} -\bs_{km'}w_{km})
	  \nonumber\\&&
			+ \frac{1}{2}(\partial_x\bar\bB)\cdot\partial_p
				(\bs_{mk}w_{m'k}+\bs_{km'}w_{km})
\ees
having the form of a Boltzmann equation, lacking the collision integral
and augmented by a coherent spin-dynamics, cf., e.g., \cite{FuchsEtAl08} and
references therein. On the r.h.s., the four terms 
describe diffusion, spin-independent acceleration, coherent spin rotation, and
spin-dependent acceleration, respectively. It is instructive to 
introduce the real-valued density and spin Wigner functions
    \bes
	w_0(x,p)&\equiv& e_{m'm}w_{m'm}(x,p),
	\nonumber\\
	{\bm w}(x,p)&\equiv&\bs_{m'm}w_{m'm}(x,p).
    \ees
Their time evolution is determined by (e.g.\ Ref.\ \cite{FuchsEtAl08})
	\bes\label{eq:boltzmann2}
	{\dot w}_0 &=&\big( -p\partial_x + x\partial_p
			+ (\partial_x V_\text{mf})\partial_p\big) w_0 
	\nonumber\\&&				
	    + \,\big(\partial_x\bB+\partial_x\bB_\text{mf}\big)\cdot\partial_p
\bw,
	\nonumber\\
	{\dot\bw} &=& \big(-p\partial_x + x\partial_p
				+ (\partial_x  V_\text{mf})\partial_p
				+  (\bB+\bB_\text{mf})\times \big) \bw
	\nonumber\\&&
	+\,\frac{1}{4}\big(\partial_x\bB+\partial_x\bB_{\text{mf}}\big)
		  \partial_p w_0,	
	\ees
where $\bB\ne0$ only during the preparation while during the time evolution to
be simulated  $\bB=0$. 

The quantum and fermionic nature of the system enters into the equations of
motion (\ref{eq:boltzmann2}) in different ways: through the initial state
$(w_0,\bw)$, via the coherent spin dynamics, and with the structure of the
mean-field
interaction (\ref{eq:mf}) stemming from the projection on spin-singlet
scattering (\ref{eq:projector}) for Fermi-statistics.
Equations~(\ref{eq:boltzmann2}) describe the collisionless regime of weak
interaction (implicitely assumed when introducing the mean-field
approximation \cite{FuchsEtAl08}). 
The collisionless regime is opposed to the hydrodynamic regime where collisions
constantly restore a local equilibrium of the momentum distribution such that
the state is described by density and velocity fields depending on $x$ only.
Both regimes can be found in the very same system:
The one-dimensional collisionless description by the set of equations
(\ref{eq:boltzmann2}) is also valid when the transversal confinement
$\omega_\perp$ is not tight enough to freeze out transversal motion completely
(as assumed here), but still tight enough to ensure quick equilibration along
the transversal directions instead \cite{DuEtAl,DuEtAlTheo}. The results
presented in this paper are, therefore, also valid in this quasi-one-dimensional
regime. The regime in between the collisionless and
hydrodynamic limit is captured by adding a collision integral to
Eqs.~(\ref{eq:boltzmann2}) tending to restore thermal equilibrium on a scale
$\tau_\text{coll}$. Considering the effect of collisions along the longitudinal
direction $x$ becomes necessary when considering stronger interactions and
longer times scales as we do. For such a regime, it has been proposed to observe
the spin-wave instability predicted by Castaing in a quantum gas
\cite{FuchsEtAl02}. 

\subsection{Initial off-equilibrium state}

Initially, the system is prepared in its spin-polarized equilibrium state,
with the spins pointing in $x$-direction, temperature $T$,
and chemical potential $\mu$. One has
	\be \label{eq:w0}
	w_0(x,p)
		=\frac{1}{2\pi}\Big\{\exp\Big[\frac{1}{T}\Big(\frac{1}{2}p^2
			+\frac{1}{2}x^2-\mu\Big)\Big]+1\Big\}^{-1}
	\ee
and $\bw=(1,0,0)^\text{t} w_0/2$ (with), or $w_{m'm}=w_0/2$. 
The zero-temperature
chemical potential, the Fermi energy for just one spin-state, simply reads $E_F=N$ with
total particle number $N=\int\!\rd x\, n_0(x)=\int\!\rd x\!\int\!\rd p\,
w_0(x,p)$. This description of the initial
state is accurate, since the spin-polarized gas is non-interacting and
the semiclassical approximation exact for a harmonic trap. In a next step, at
time $t=0$, during a short preparatory pulse a $z$-polarized magnetic field
gradient is applied, captured by
$\bB(x)=qx\delta(t){\bm e_z}$. 
A spin spiral of wave length $\lambda_s=2\pi/q$ is created,
while $w_0(x,p)$ is still given by Eq.~(\ref{eq:w0}) one has
	\be\label{eq:spiral}
	\bw(x,p) = \big(\cos(qx),\sin(qx),0\big)^\text{\!t} \,\frac{w_0(x,p)}{2}.
	\ee
or $w_{m'm}(x,p)=\exp[\ri qx(m-m')]w_0(x,p)/2$. 
With a simple spin rotation, we have prepared a state far from thermal
equilibrium. Apart from (i) having created a rather artificial spiral spin
configuration (\ref{eq:spiral}) (favorable neither
with respect to energy nor entropy), we have also increased the number of
available single-particle states from one spin state to two. 
The latter has two consequences: (ii) the phase-space density configuration
(\ref{eq:w0}) is far from being thermal (for half the number of particles per
spin state
having the same kinetic energy as before, a thermal distribution is
characterized
by a lower chemical potential and a higher temperature), and (iii) we have suddenly
introduced interaction to the system. The
combination of (i) and (iii) will lead to robust dynamical spin segregation. 

\subsection{Semiclassical versus mean-field dynamics}
When integrating the time evolution for many fermions, the semiclassical
phase-space equations~(\ref{eq:boltzmann2}) are usually much easier to treat
numerically than the Hartree-Fock mean-field equations.~(\ref{eq:motion}), even though
the interaction is non-local in the former,  
    \bes\label{eq:projection}
      V_\text{mf}(x)&=&\frac{g}{2}\int\!\rd p\,w_0(x,p),\quad
      \nonumber\\
      \bB_\text{mf}(x)&=&-2g\int\!\rd p\,\bw(x,p).
      \ees 
This is exemplified by our state (\ref{eq:w0}) and (\ref{eq:spiral}): In phase space, it 
has a linear extent $\Delta\sim\sqrt{\max(\mu,T)}$ [we define
$\Delta\equiv 2\max(\sqrt{2N},2\sqrt{T})$], while it varies on the scale
$\delta \sim \min\big(T/\Delta, \lambda_s\big)$ stemming either from the former
equilibrium [roughly estimating $\frac{\rd w_0}{\rd \varepsilon}\frac{\rd
\varepsilon}{\rd s}\sim \frac{w_0}{T}\Delta$ with
$\varepsilon=\frac{1}{2}(x^2+p^2)$ and $s=x,p$] or from the induced spin spiral.
For a reasonable phase-space representation
one, thus, requires a grid of just more than $(\Delta/\delta)^2\sim N$ points. On the 
other hand, the real-space single-particle density matrix $n_{m'm}(x',x)$
varies on the much shorter length $\delta'\sim\Delta^{-1}$ in each argument,
calling for more than $\Delta^4\gtrsim N^2$ grid points.

\section{\label{sec:simulation} Simulation of dynamics}

\subsection{\label{sec:parameters}System parameters}
In the following we consider Li$^6$ atoms with mass $M\approx 1.0\cdot10^{-26}$ kg
and the scattering length tuned down to $a_s\approx2.4\cdot10^{-10}{\rm m}$ by using a 
magnetic Feshbach resonance; the trap frequencies read $\omega=2\pi\cdot60$ Hz and
$\omega_\perp=2\pi\cdot3.6$ kHz. Returning to dimensionless units, we obtain
the weak coupling $g\approx0.055$.
The initial spin-polarized equilibrium is characterized by the
particle number $N=E_F=100$ and by the temperature $T$ taking values of either
$T/E_F=$ 0.2, 1, or 5 corresponding to the degenerate, intermediate and non-degenerate regime.
According to these values one finds: the chemical potentials $\mu/E_F\approx$ 1.0, 0.54, -7.2; cloud
extensions $\Delta\approx$ 28, 40, 89; maximum densities $n_0(0)\approx$ 4.4, 3.3, 1.8; and maximum
mean-field potential strengths $n_0(0)g\approx$ 0.24, 0.18, 0.097 that are small compared to the trap
frequency, being 1 in our units, and extremely small with respect to typical single-particle energies
$\max(T,\mu)\gtrsim 100$. 
In addition to the temperature, we also vary the
spiral wave length $\lambda_s$ and consider either $\Delta/\lambda_s=$ 1, 2 or 5
windings of the spin spiral within the atom cloud. For these
conditions, we integrate the time
evolution using a MacCormack method \cite{MacCormack03} and trust results that
coincide for grid-sizes
$300^2$ and $600^2$ for a phase space region of linear extension $\approx2\Delta$.

\subsection{Observation of spin segregation}

\begin{figure}[t]\centering
\includegraphics[width = 1\linewidth]{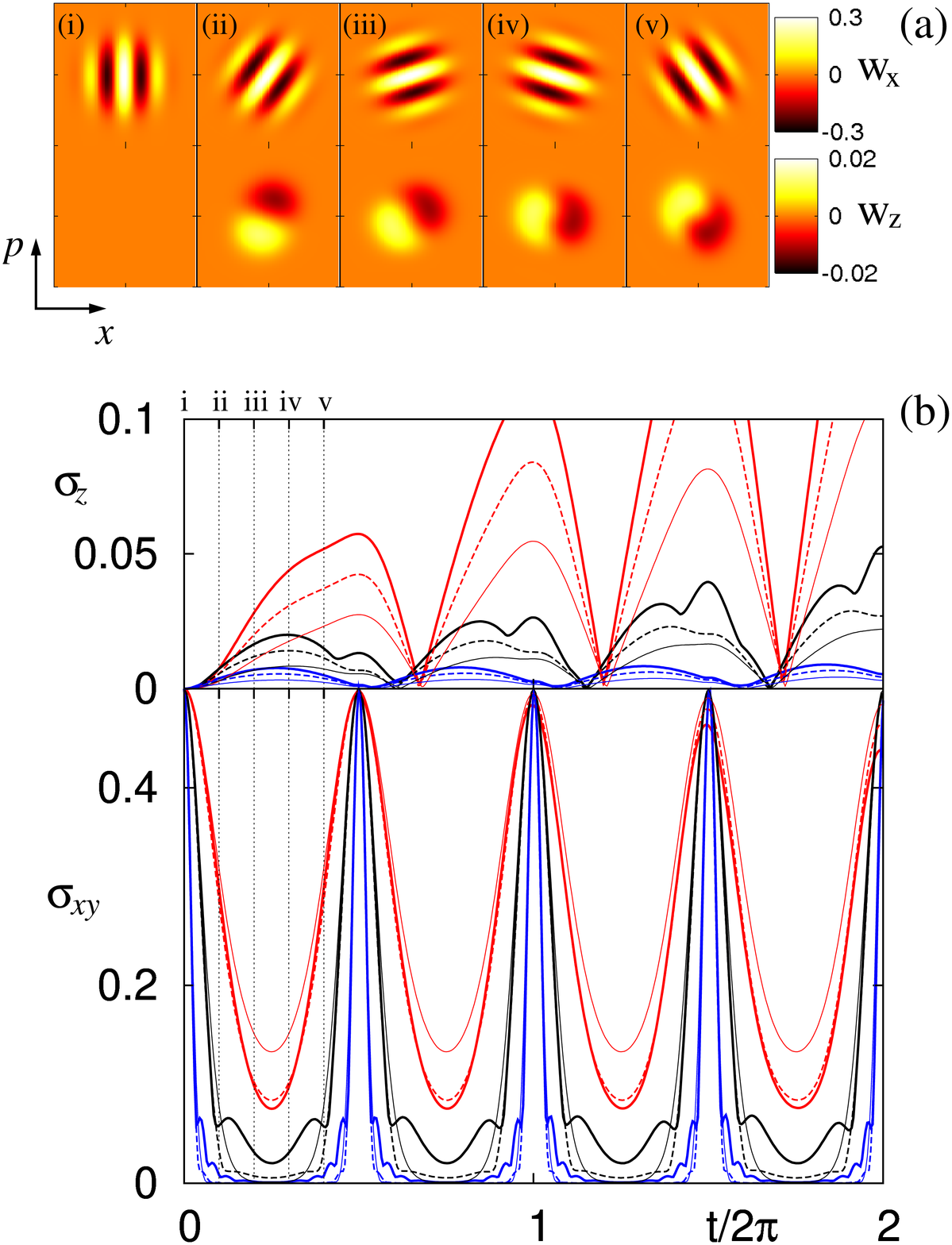}
\caption{\label{fig:short} (Color online) 
(a) Wigner functions $w_x(x,p)$ and $w_z(x,p)$ at five times [i-v,
as indicated in (b)] during the first half cycle; $T/E_F=1$,
$\lambda_s/\Delta=0.5$, both $x$ and $p$ range from -40 to 40. 
The motion in phase space is governed by an overall rotation at the trap
frequency, during which the
$z$-component of the Wigner function slowly builds up two domains. 
(b) Time evolution of the averaged spatial polarizations $\sigma_z$ and $\sigma_{xy}$ for
$T/E_F=$ 0.2, 1, 5 (thick solid, dashed, and thin solid lines, respectively) and for
$\lambda_s/\Delta=$ 1, 0.5, 0.2  (upper triple of red, central triple of black, and lower 
triple of blue lines, respectively). While for the shorter wave lengths $\lambda_s/\Delta$ 
the spatial $xy$-polarization shows rapid collapses followed by periodic revivals, the emerging
spatial $z$-polarization undergoes smooth oscillations as a signature of the
formation of only two domains in phase space.} 
\end{figure}

On a short time scale $\sim 1$, the system evolves mainly as determined by the harmonic trapping
potential. That is, neglecting interaction completely for the moment, 
$\big(w_0(x,p),\bw(x,p)\big)$ simply rotates in phase space at constant angular velocity 1 (in
units of the trap); each point of the Wigner function follows a classical circular orbit. This 
behavior can be observed clearly in the first row of Fig.~\ref{fig:short}(a) showing the evolution
of $w_x$ during half a cycle. This simple dynamics in phase space translates into a more
involved evolution of the spatial polarization $\bn(x)$, obtained by projecting
$\bw(x,p)$ onto the $x$-axis, $\bn(x)=\int\!\rd p\,\bw(x,p)$.
Fig.~\ref{fig:short}(b)
shows the averaged absolute spatial $xy$-polarization, 
  \be
    \sigma_{xy}\equiv\frac{1}{N}\int\!\rd x\,[n_x^2(x)+n_y^2(x)]^{\!1/2},
   \ee               
during the first two cycles. A rapid collapse of $\sigma_{xy}$ followed by periodic revivals can be observed, the more
pronounced the larger the number of spiral windings $\Delta/\lambda_s$ in the atom cloud.

\begin{figure}[t]\centering
\includegraphics[width = 1\linewidth]{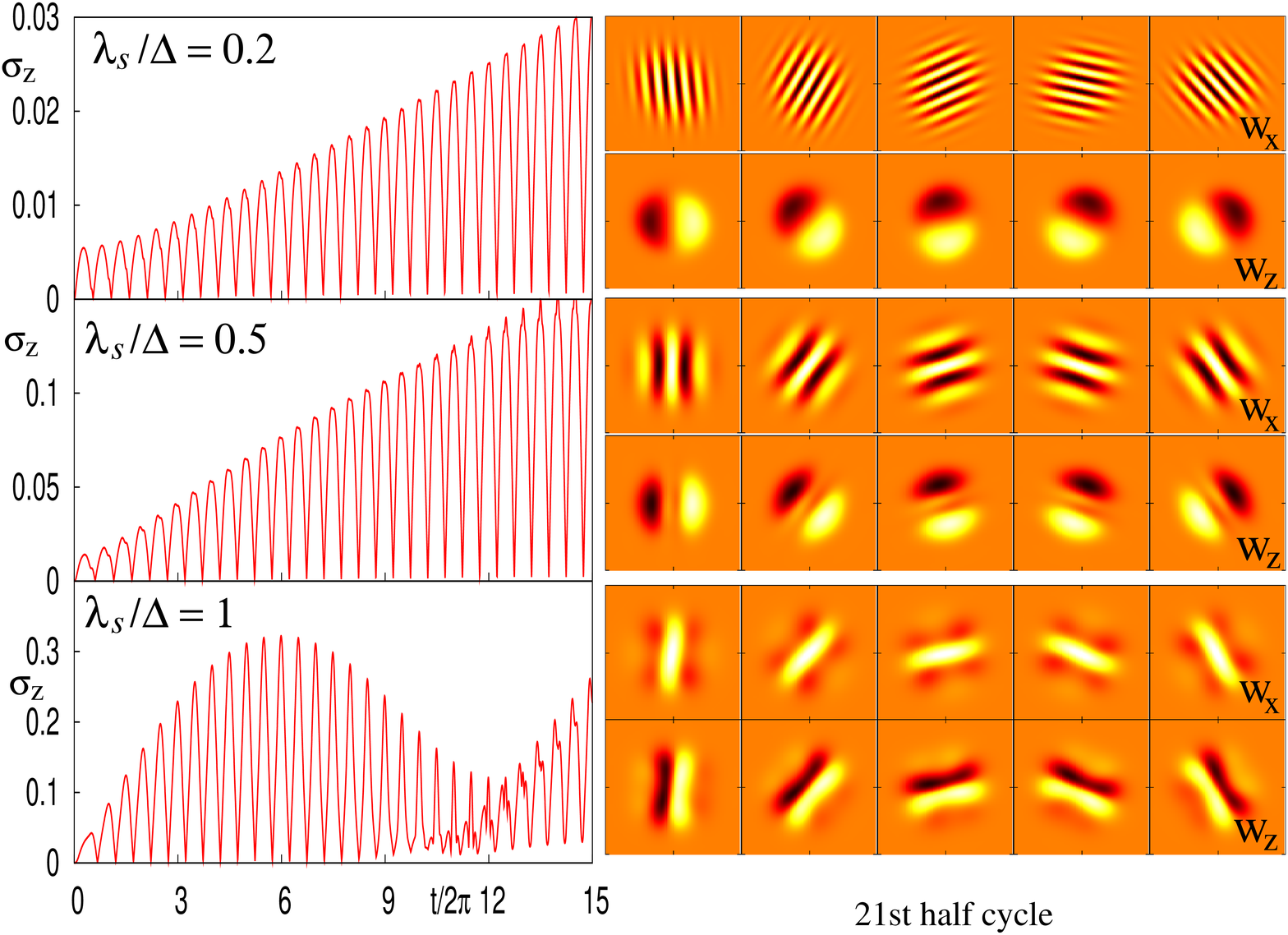}
\caption{\label{fig:long} (Color online) Evolution on longer times, for $T/E_F=1$, and different
wave lenghts $\lambda_s/\Delta$. The l.h.s.\ shows the averaged
spatial absolute $z$-polarization, $\sigma_z$, versus time. On the
r.h.s.\ the Wigner functions $w_x$ and $w_z$ are plotted for five instants of times like in
Fig.~\ref{fig:short}(a), but during the 21st half cycle (arbitry color scale). 
Initially $\sigma_z$ increases linearly in time with a rate that
is controlled by the wave length $\lambda_s/\Delta$. Irrespective of the number of windings
$\Delta/\lambda_s$ (directly visible in $w_x$), $w_z$ developeds \emph{two}
oppositely polarized domains.} 
\end{figure}

\begin{figure}[t]\centering
\includegraphics[width = 1\linewidth]{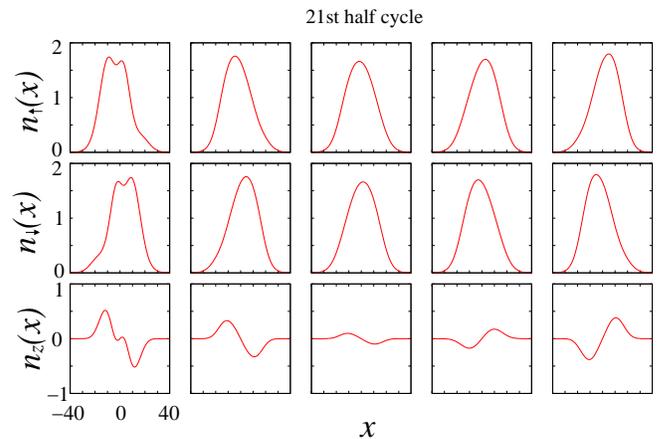}
\caption{\label{fig:densities} (Color online) Spatial densities $n_\up(x)$ and $n_\dw(x)$
as well as $z$-polarization
$n_z(x)=\frac{1}{2}[n_\up(x) - n_\dw(x)]$ at five instants of time during the 21st half cycle,
for $T/E_F=1$ and $\lambda_s/\Delta=0.2$. Times and parameters correspond to the middle row of the
r.h.s.\ of Fig.~\ref{fig:long}. Spin $\up$ and $\dw$ particles segregate and perform phase-opposed
dipole oscillations in the trap.}
\end{figure}

During a single cycle of oscillation in the trap the weak interaction causes
only small deviations from a simple rotation in phase space. The small
modification of the trap frequency and the slight anharmonicities
caused by the scalar part of the mean-field potential $V_\text{mf}(x)$ are
hardly noticeable.
However, the effect of interaction becomes apparent in $w_z$, being
zero initially [Fig.~\ref{fig:short}(a), note the different color scales for
$w_x$ and $w_z$]. Though still small, $w_z$ develops a characteristic pattern
induced by the magnetic mean field
$\bB_\text{mf}$. Namely, in $w_z$ two domains of opposite $z$-polarization form.
This spin segregation in phase space corresponds to phase-opposed dipole oscillations of the
$\up$- and the $\dw$-polarized domain in the trap (see also Fig.~\ref{fig:densities}). Remarkably,
the spin segregation does not reproduce the structure of the initially created
spin spiral of wave length $\lambda_s$. 
The formation of two spin domains (and only two) is a very robust effect; we
find it for all
pitches of the spin spiral considered here. The upper panel of Fig~\ref{fig:short}(b) shows the
averaged absolute spatial $z$-polarization 
    \be
      \sigma_z\equiv\frac{1}{N}\int\!\rd x\,[n_z^2(x)]^{\!1/2}.
    \ee 
As a consequence of spin segregation in phase space, $\sigma_z$ oscillates in time, however,
because only two domains are formed, it does not feature sharp collapses like
$\sigma_{xy}$ does for small $\lambda_s/\Delta$.

In Fig~\ref{fig:long} we present data for longer times, for the
original temperature $T/E_F=1$ and for three different spiral
wave lengths $\lambda_s/\Delta$: From cycle to cycle the spin segregation becomes more and more
pronounced as visible from $\sigma_z$ and from the snapshots on the r.h.s.\
showing $w_z$ during the
21st half cycle. The rotation in phase space of the two oppositely polarized domains corresponds
to phase-opposed dipole oscillation of $\up$ and $\dw$ spins in the trap. This behavior is visible 
in Fig.~\ref{fig:densities} showing the spatial densities $n_\up(x)$ and $n_\dw(x)$ as well as the
spatial polarization $n_z(x)=\frac{1}{2}[n_\up(x) - n_\dw(x)]$ during the 21st cycle.
While a $z$-polarization builds up, the spiral spin structure in the $xy$-plane decreases but
stays intact (cf.\ $w_x$ during the 21st half cycle shown in Fig.~\ref{fig:long}).
The total density Wigner function $w_0$ hardly changes during
the time-evolution also for longer times (not shown). The spin segregation can directly be
controlled by the number of windings $\Delta/\lambda_s$ of the spin spiral within the cloud,
the more windings the slower the segregation builds up. The fastest
segregation is observed for $\Delta/\lambda_s=1$, here already after 10
half cycles deviations from a linear
increase is found and a more complex dynamics sets in (Fig.~\ref{fig:long}). 

\section{\label{sec:effective} Explanation by dynamically induced
long-range interaction}

\subsection{\label{sec:eff}Effective Description}

In order to give an intuitive explanation for the spin segregation, 
let us describe the system in the rotating phase-space frame with the new
coordinates $x'=x\cos(t)-p\sin(t)$ and $p'=p\cos(t)+x\sin(t)$ describing
classical orbits in the trap. In that frame 
$\bw'(x',p',t)\equiv\bw(x(x',p',t),p(x',p',t),t)$ is
stationary for vanishing interaction. However, interaction, represented by the mean-field potentials
(\ref{eq:projection}), is now time-dependent, since it is obtaned by projecting
onto the $x$-axis that rotates with respect to the new frame. For example, the
magnetic mean field $\bB_\text{mf}'(x',p',t) = \bB_\text{mf}(x(x',p',t),t)$
reads
      \be\label{eq:B'}
	\bB_\text{mf}'(x',p',t)=
	    -2g\int_{-\infty}^\infty\!\rd s\, 
		\bw'\big(x'-s\sin(t),p'+s\cos(t),t\big).
      \ee
The time-dependence of the mean field $\bB_\text{mf}'$ originates on the
one hand from the rotation of the integration axis at trap frequency and on
the other hand from
the time-dependence of the Wigner function $\bw'(x',p',t)$. In the rotating
frame, the latter is solely governed by the weak interaction and slow compared
to the oscillation in the trap. We can use this difference in time scales to
separate the dynamics on short times from that on longer times.
We assume that a single oscillation in the trap is not affected by
the weak interaction. This allows us, in turn, to integrate out the rapid
oscillations in the trap when studying the dynamics on longer times where
interaction does play a role;  we approximate
     \bes
	&&\bB_\text{mf}'(x',p',t)
	  \approx\bB^{\text{eff}}_\text{mf}(x',p',t)
	\\\nonumber
	  &&\equiv\frac{-2g}{2\pi} \int_0^{2\pi}\!\rd\tau
	 \int_{-\infty}^\infty\!\rd s\, 
	\bw'\big(x'-s\sin(\tau),p'+s\cos(\tau),t\big)
      \ees
giving
	\be  \label{eq:Bcyc}
	  \bB^{\text{eff}}_\text{mf}(x',p',t)  = 
	    \int\!\rd\tilde{p}\int\!\rd\tilde{x} 
	    \frac{-2g/\pi}{\sqrt{\tilde{x}^2+\tilde{p}^2}}
	      \bw'\big(x'+\tilde{x},p'+\tilde{p},t\big).
	\ee
By averaging over a cycle, we have obtained an effective mean-field potential
that corresponds to a time-independent isotropically long-ranged interaction in
phase space. By oscillating in the trap, the system
dynamically acquires a spatially long-range interaction.

\subsection{Zero-order semiclassical mean-field interaction}
We can simplify the description further, again arguing that interaction is
weak and the mean-field potential small compared to the trap. For the
mean-field contribution $V^\text{mf}_{m'm}$ of the potential $\bar
V_{m'm}(x)=V^\text{mf}_{m'm}(x)+\frac{1}{2}x^2$, appearing in the infinite
series on the r.h.s.\ of the equation of motion (\ref{eq:series}), we truncate
the series already after $\alpha=0$ instead of $\alpha=1$. In the set of
equations (\ref{eq:boltzmann2}), this approximation
corresponds to neglecting the mean-field-induced acceleration by
dropping terms involving the gradients $\partial_x\bB_\text{mf}$ and
$\partial_xV_\text{mf}$. We keep, however, the spin-rotating
term $\bB_\text{mf}\times\bw$ stemming from the order $\alpha=0$.
Together with cycle averaging, in the rotating phase-space frame, we arrive
at the effective equations of motion
      \bes\label{eq:simple}
	\dot w_0'(x',p',t) &=& 0
	\nonumber\\
	\dot\bw'(x',p',t) &=&  
	  \bB^{\text{eff}}_\text{mf}(x',p',t)\times\bw'(x',p',t).
      \ees
The second equation describes the time evolution of the
polarization field $\bw'$ in the $x'p'$-plane. At each point the polarization
rotates in the mean field $\bB^{\text{eff}}_\text{mf}$ such that $|\bw'|$ stays
constant.

\subsection{Growth of $z$-polarization}
The spin segregation observed numerically can now be explained by
first-order time-dependent perturbation theory, predicting according to
Eqs.~(\ref{eq:simple}) initially a linear growth of the z-polarization, 
    \be\label{eq:linear}
    w_z(x',p',t) \simeq  
[\bB^{\text{eff}}_\text{mf}(x',p',0)\times\bw'(x',p',0)]_z t,
    \ee 
as we can observe it on the l.h.s.~of Fig.~\ref{fig:long}. Deviations
from the linear growth (\ref{eq:linear}) appear as soon as $w_z'$
becomes comparable to $|\bw'|$ as visible in the lower left plot of
Fig.~\ref{fig:long}.
Figure~\ref{fig:pert} shows the rate
$[\bB^{\text{eff}}_\text{mf}(0)\times\bw'(0)]_z$ computed for the intermediate
temperature $T/E_F=1$ and different spiral wave lengths $\lambda_s/\Delta$. 
Notably $[\bB^{\text{eff}}_\text{mf}(0)\times\bw'(0)]_z$ always shows a
pattern with two oppositely polarized domains, irrespective of the number
of windings $\Delta/\lambda_s$ \footnote{%
Slight deviations from this behavior are found for the low temperature
$T/\mu=0.2$ where the phase-space density profile has a
step-like behavior.}. This explains the previously observed
segregation of $\up$ and $\dw$ polarization. The formation of two domains only
can be understood as follows:

\begin{figure}[t]\centering
\includegraphics[width = 0.6\linewidth]{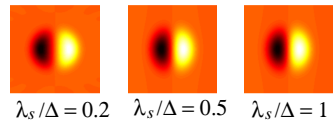}
\caption{\label{fig:pert} (Color online) Cycle-averaged rate of the creation of
$z$-polarization computed for the initial state,
$[\bB^{\text{eff}}_\text{mf}(0)\times\bw'(0)]_z$, for $T/E_F=1$ and
different spiral wave
lenghts $\lambda_s/\Delta$ (arbitrary color scale). It explains the formation of two oppositely
polarized domains in $w_z$ that is visible in Fig.~\ref{fig:long}.}  
\end{figure}

According to Eq.~(\ref{eq:Bcyc}), the spin polarization $\bw'$ at a given
point $(x',p')$ in phase space feels a magnetic mean field that mainly depends
on the polarization found in phase-space areas close by. Within
the
phase-space neighborhood of $(x',p')$, in turn, phase-space areas showing the
largest polarization perpendicular to $\bw'(x',p')$ contribute
most. Since for the initial state $|\bw'|=w_0'/2$ increases towards the origin
$x'=p'=0$, at a given point $(x',p')$ the effective magnetic mean field
$\bB^{\text{eff}}_\text{mf}$ is dominated by the polarization found when
slightly moving along the direction of the spiral towards the origin.
On one side of the spiral this results always in the creation of a positive
$z$-polarization, on the other side always to the creation of a negative
$z$-polarization. This explains the creation of two domains. Moreover, the
strength of the local mean field depends on the spatial variation of the spin
density $|\bw'|$ compared to the spiral wave length $\lambda_s$; the
smaller $\lambda_s/\Delta$ the slower the spin segregation, as observed
numerically in Fig.~(\ref{fig:long}).  

We have identified the mechanism underlying the observed spin segregation.
Obviously, the phenomenon does not depend on the sign of the spin-spin coupling.
We have checked numerically that it is equally observable for attractive
interaction, giving reversed polarizations. Moreover, it can also be expected
for non-condensed ``spin-1/2'' bosons which are equally described by
Eqs.~(\ref{eq:boltzmann2}), but with the exchange interaction giving rise to a
spin-coupling of opposite sign. 

\section{Experimental signatures}

One can measure the dynamical spin segregation by state-sensitive absorption imaging either 
\emph{in situ} (as in the experiment by Du et al.~\cite{DuEtAl}) or after a time of flight. In the
latter case one can also use Stern-Gerlach separation to distinguish $\up$ and $\dw$ particles.
An \emph{in situ} measurement gives the spatial distributions $n_m(x)=\int\!\rd p\,w_{mm}(x,p)$ of
both spin states $m=\up,\dw$; the images after a time of flight reveal their momentum distribution
$\tilde{n}_m(p)=\int\!\rd x\,w_{mm}(x,p)$. One can then determine the $z$-polarization in space,
$n_z(x)=\frac{1}{2}[n_\up(x)-n_\dw(x)]$, and momentum,
$\tilde{n}_z(p)=\frac{1}{2}[\tilde{n}_\up(p)-\tilde{n}_\dw(p)]$. As shown in Fig.~\ref{fig:densities},
the dynamical spin segregation corresponds to phase-opposed dipole oscillations
of $\up$ and $\dw$ spins in space. The momentum distributions will show the very
same behavior, but shifted in time by the quarter of a
cycle (because momentum densities are obtained by projecting
the Wigner function onto the $p$-axis).

\section{Conclusions}

The phenomenon of dynamical spin segregation predicted here is different from the effect observed by
Du et al.\ described in Refs.~\cite{DuEtAl,DuEtAlTheo}. In their case no spin spiral is created
initially, instead an inhomogeneous external magnetic field is present throughout, leading eventually
to a spherical symmetric spin segregation in phase space between an inner core and an outer shell. 
The phenomenon described here also differs from the physics of the spin-wave instability investigated
by Conduit and Altman \cite{ConduitAltman10}. They consider the same spiral spin structure as 
initial state, but strong repulsive interaction. In contrast to the dynamical spin segregation into
two counter oscillating domains found here, crucially depending on the presence of the trap, their
spin-wave instability leads to spatial (non oscillatory) domain formation, not requiring a trap and 
with the domain size controlled by the spiral wave length.

The system's dynamics described here can be called \emph{self driven}. The
transformation to the
co-rotating frame in phase space corresponds to the transformation to the Dirac picture on the full 
quantum many-body level, where the quadratic Hamiltonian $\Ho_0=\Ho-\Ho_\text{int}$ constitutes the
unperturbed  problem. In the Dirac picture, the time evolution is generated solely by the
time-dependent interaction Hamiltonian
$\Ho^\text{d}_\text{int}(t)=\exp(-\ri\Ho_0t/\hbar)\Ho_\text{int}\exp(\ri\Ho_0t/\hbar)$.
Thanks to the equidistant ladder spectrum of the harmonic trap, it is time periodic, 
$\Ho^\text{d}_\text{int}(t+\mathcal{T})=\Ho^\text{d}_\text{int}(t)$ with
$\mathcal{T}=2\pi/\omega$, like the Hamiltonian of a driven system. This
additional symmetry has strong consequences for the dynamics. It
allows us to find a time-independent effective description 
$\Ho_\text{eff}=\frac{1}{\mathcal{T}}\int_0^\mathcal{T}\!\rd t\,
\Ho^\text{d}_\text{int}(t)$ for the dynamics on longer time scales that does not
depend on the details of the short-time dynamics. For the system described here,
the effective description, in form of the mean field potential
(\ref{eq:Bcyc}), contains a spatially long-range interaction (isotropic in phase
space) that the original Hamiltonian did not possess and that explains the
spin segregation observed. The system has dynamically acquired novel properties.
A similar situation is found for example for interacting particles in tilted
lattice systems with the single-particle spectrum given by the
Wannier-Stark ladder \cite{KolovskyBuchleitner03}.
The separation of time scales found here resembles also the physics of driven
many-body systems as it has been studied in lattice systems subjected to
off-resonant external driving. These systems are equally described  by an
approximate effective time-independent Hamiltonian on long
times~\cite{EckardtEtAl}. 

The dynamical spin segregation is -- like itinerant ferromagnetism -- caused by
exchange interaction. However, while the Stoner transition to a ferromagnetic phase is an
equilibrium effect requiring fairly strong interaction (as well as spatial dimensionalities 
larger than one), the robust effect described here happens far from equilibrium and does not
need strong interaction but, instead, sufficiently long times to build up. 

We gratefully acknowledge discussion with Luis Santos and support by the Spanish MICINN
(FIS 2008-00784, FPI-fellowship), the A.v.~Humboldt foundation, ERC Grant QUAGATUA, and
EU STREP NAMEQUAM.


\end{document}